# Phonon heat transport with squeezing-based symmetry breaking

Yan Cao[1,†], Cheng Yang[1,†], Xintong Gu[1], Shenzhu Wang[1], Jiteng Sheng[1,2,*], and Haibin Wu[1,2,3,4,*]

[1]Institute of Quantum Science and Precision Measurement, School of Physics, East China Normal University, Shanghai 200062, China

[2]Collaborative Innovation Center of Extreme Optics, Shanxi University, Taiyuan 030006, China

[3]Shanghai Branch, Hefei National Laboratory, Shanghai 201315, China

[4]Shanghai Research Center for Quantum Sciences, Shanghai 201315, China

†*These authors contributed equally to this work.*

**jtsheng@lps.ecnu.edu.cn; hbwu@phy.ecnu.edu.cn*

**The controllability of phonon thermal transport is fundamental for numerous technologies, from cooling high-performance chips to managing heat in quantum computing. Despite extensive efforts, on-demand, real-time control of phonon heat flow remains elusive, being fundamentally constrained by the temperature gradient. Here we achieve this long-sought phonon heat transport by introducing a novel mechanism using quantum squeezing in a cavity optomechanical system. We find that phonon squeezing of near-ground-state mechanical resonators via an optomechanically induced parametric process breaks the continuous U(1) rotation symmetry in phase space, and consequently the symmetry in the heat current structure. We reveal that heat flow under a constant temperature gradient can be deterministically amplified, reduced, or even reversed, a capability previously unattainable. With phonon squeezing, we achieve over twentyfold amplification of heat flow and reversal within 30 milliseconds. Importantly, quantum discord analysis reveals a direct connection between heat flow reversal and quantum correlation. Our results establish quantum squeezing as a versatile platform for on-demand phononic thermal control, opening avenues for active heat management in quantum devices and thermal logic circuits.**

Phonons, the quanta of mechanical vibrations, are the fundamental heat carriers. Controlling phonon heat flow has thus long been a pursued yet challenging goal in fundamental physics and nanoscience [1-5]. In stark contrast to the sophisticated control achieved for electrons and photons, dynamic control of phonon thermal transport lags far behind. This disparity arises from fundamental obstacles: the second law of thermodynamics imposes a time arrow on heat flow [6], and neutral phonons exhibit negligible direct coupling to external fields [7].

Over the past decades, substantial theoretical and experimental efforts have focused on developing phononic thermal control devices that exploit structural asymmetry, nonlinearity, and phase-transition materials to achieve thermal rectification, switching, and logic operations [8-15]. External stimuli including electric/magnetic fields, strain, pressure, and light excitation have also been explored for active thermal manipulation [16-24]. As novel schemes for thermal transport, field-mediated phonon heat transfer has been demonstrated via short-range Casimir force [25] and long-range optomechanical interactions [26]. Despite such progress, on-demand phonon heat flow with large tunability (flow direction) and fast response at quantum-level operation remains elusive.

Here, we report the first experimental realization of on-demand, real-time control of phonon heat flow via quantum squeezing in a cavity optomechanical system with two near-ground-state cooled nanomechanical resonators. Quantum squeezing has recently emerged as a powerful tool for achieving nonreciprocal transport [27-29] and enhanced quantum effects [30-33]. However, phonon heat transport with quantum squeezing is lacking. We demonstrate that the phonon heat current, under a constant temperature gradient, can be deterministically amplified, suppressed, or even fully reversed by introducing phonon squeezing to one resonator. This exceptional controllability arises from quantum squeezing breaking the continuous U(1) rotational symmetry of the phase space, which in turn breaks the symmetry of the heat current components that stem from cross-quadrature correlations of mechanical motions. Experimentally, we achieve over twentyfold amplification and complete reversal of heat flow within 30 milliseconds. Furthermore, we establish a direct link between heat current reversal and quantum discord, with the minimum of quantum discord coinciding with the zero point of one heat current component. Our results establish quantum squeezing as a versatile resource for on-demand thermal manipulation and reveal an intrinsic physical connection between quantum correlation and phonon heat transfer.

To illustrate the quantum-squeezing-engineered phonon heat transport, we consider a system of two coupled mechanical resonators, $M_1$ and $M_2$ (Fig. 1a). Both resonators as phonon modes are initially cooled close to their motional ground states and coupled to thermal reservoirs at distinct temperatures, i.e. $M_1$ to a cold bath and $M_2$ to a hot bath. The system can be described by an effective Hamiltonian (see Supplementary Note 1)

$$H_{\text{eff}} = i\frac{\Gamma_{\text{par}}}{4}\left(\hat{b}_n^{\,2} - \hat{b}_n^{\dagger 2}\right) + \frac{\eta}{2}\left(\hat{b}_1\hat{b}_2^{\dagger} + \hat{b}_1^{\dagger}\hat{b}_2\right). \quad (1)$$

Here $\hat{b}_{1,2}$ and $\hat{b}_{1,2}^{\dagger}$ denote the annihilation and creation operators for the phonon modes. The first term represents the squeezing Hamiltonian with strength $\Gamma_{\text{par}}$, where n=1(2) specifies that the squeezing is applied exclusively to $M_1$ ($M_2$). The second term describes the coherent coupling between the phonon modes with strength η due to optomechanical interaction [26]. In the absence of squeezing ($\Gamma_{\text{par}} = 0$), $H_{\text{eff}}$ possesses a continuous U(1) rotation symmetry. A finite squeezing establishes a preferred axis in phase space and explicitly breaks the continuous U(1) symmetry, thereby reducing it to $Z_2$ symmetry (invariant under the transformation $\hat{b}_{1,2} \to -\hat{b}_{1,2}$ ). This symmetry reduction reconfigures the phononic interactions and enables squeezing-engineered phonon heat transport.

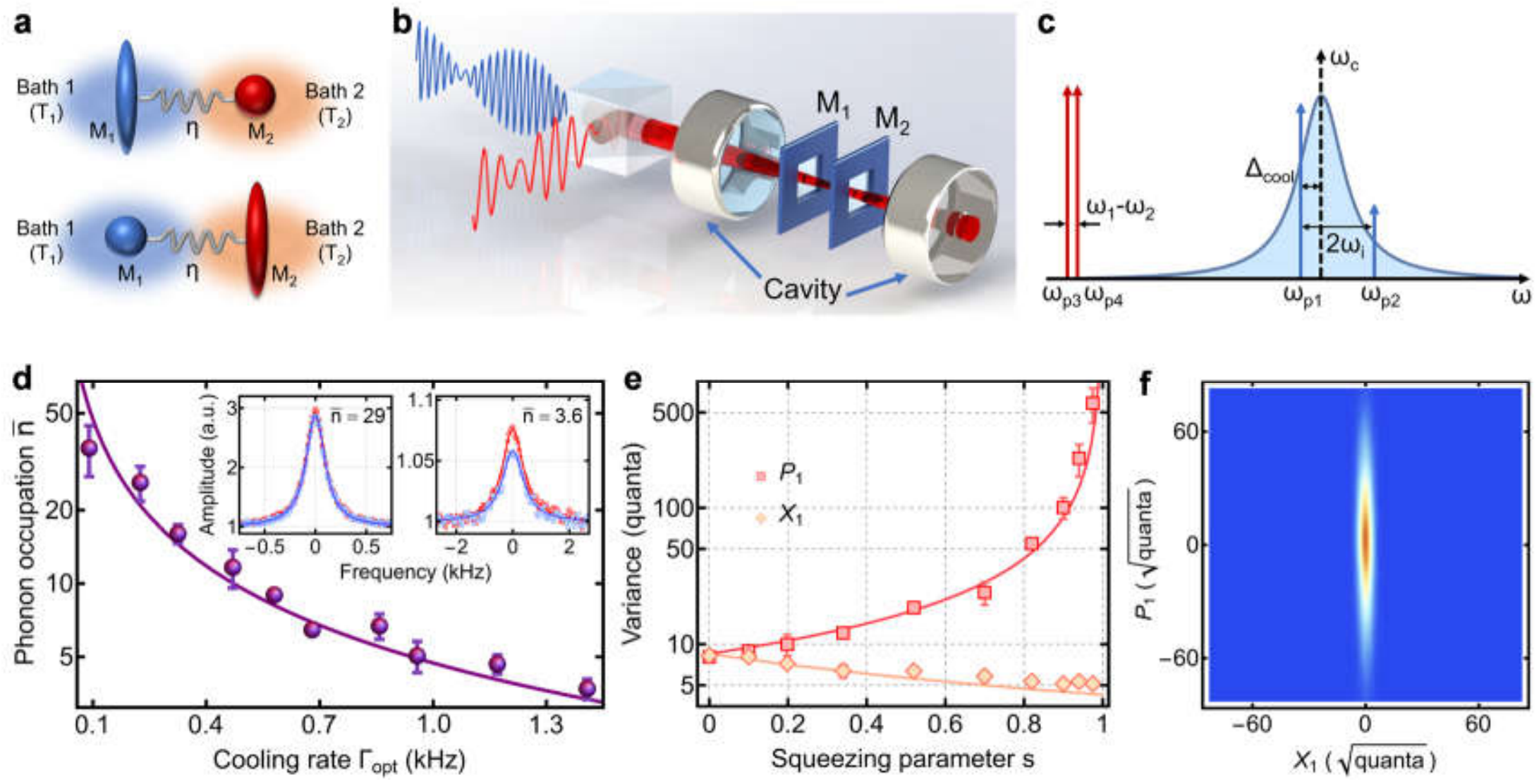

**Fig. 1. Experimental setup and characterization of the two-membrane cavity optomechanical system for quantum-squeezing-engineered phonon heat transport.** **a**, Two coupled mechanical oscillators $M_1$ and $M_2$ (coupling η), each linked to a separate thermal bath ($T_1$: low temperature, $T_2$: high temperature; shaded areas), with phonon squeezing (ellipse) applied to $M_1$ or $M_2$. **b**, Schematic of the two-membrane cavity optomechanical system, where two nanomechanical membranes ($M_1$ and $M_2$) are placed inside an optical cavity pumped by two pairs of beating lasers (red and blue wavy lines). **c**, Optical pump using four laser fields for simultaneous cooling, squeezing, and coherent coupling of two phonon modes. The strong red-detuned laser ($\omega_{p1}$) provides side-band cooling. $\omega_{p1}$ together with a blue-detuned laser ($\omega_{p2}$) induces phonon squeezing when $\omega_{p2}$-$\omega_{p1}$ = $2\omega_1$ (or $2\omega_2$). Two far-red-detuned fields with frequency difference $\omega_{p4}$-$\omega_{p3}$ = $\omega_1$-$\omega_2$ generate parametric coherent coupling between $M_1$ and $M_2$ for the purpose of phonon heat transport. **d**, Measured phonon occupation of the mechanical resonator as a function of the optomechanical cooling rate from the red-detuning laser $\omega_{p1}$. The circles (experiment) and solid line (simulation). Inset: heterodyne spectra of the Stokes (red) and anti-Stokes (blue) sideband amplitudes, with corresponding phonon occupations of 29 and 3.6. **e**, Measured quadrature variances of $X_1$ (orange diamonds) and $P_1$ (red squares) versus squeezing parameter s. **f**, Measured probability distributions of the squeezed mechanical resonator $M_1$ in $X_1$-$P_1$ phase space at s = 0.985.

In the experiment, we employ a two-membrane-in-the-middle cavity optomechanical system [34-36], in which two silicon nitride nanomembranes, $M_1$ and $M_2$, are spatially separated inside a high-finesse optical cavity (Fig. 1b). The entire apparatus is housed in a cryogenic vacuum chamber maintained at a base temperature of 2.7 K and a pressure of $10^{-8}$ mbar. We focus on the (3,3) vibrational eigenmodes of the two membranes, with resonance frequencies $\omega_1/2\pi$ = 1.165 MHz and $\omega_2/2\pi$ = 1.134 MHz. The optical cavity, with linewidth $\kappa/2\pi \sim 1.6$ MHz, is driven by two pairs of laser fields, with frequency detunings relative to the cavity resonance as illustrated in Fig. 1(c).

By applying a strong red-detuned laser ($\omega_{p1}$), the phonon modes are cooled from cryogenic temperatures to near their quantum ground state. Figure 1d illustrates the measured phonon occupation as a function of the cooling rate ($\Gamma_{1,2}^{\mathrm{opt}}$), which increases the effective mechanical damping ($\gamma_{1,2}^{\mathrm{eff}} = \gamma_{1,2} + \Gamma_{1,2}^{\mathrm{opt}}$), yielding a minimum average phonon number $\bar{n}_{1,2}^{\mathrm{eff}}$ of approximately 3.6. The corresponding heterodyne spectra (Fig. 1d, insets) reveal a pronounced asymmetry between the Stokes (red) and anti-Stokes (blue) sidebands, providing a clear signature of the quantum nature of the mechanical resonator [37]. To induce phonon squeezing in this near-ground-state regime, we apply a pair of beating lasers ($\omega_{p1}$ and $\omega_{p2}$) with a frequency difference precisely matched to $2\omega_1$ (or $2\omega_2$ for squeezing the second mode) [38,39]. This induces a parametric squeezing process in $M_1$, described by the first term of Eq. (1). The variances of two orthogonal mechanical quadratures ($X_1$, $P_1$) are shown in Fig. 1e as a function of the squeezing parameter s (defined as $\Gamma_{\mathrm{par}} / \gamma_{1,2}^{\mathrm{eff}}$). The phase-space probability distribution at s = 0.985 (Fig. 1f) exhibits clear squeezing in $X_1$ approaching the 3 dB limit, accompanied by corresponding antisqueezing in $P_1$ (see Supplementary Notes 2-4).

To establish the coupling channel for phonon heat transport, a second pair of beating lasers ($\omega_{p3}$ and $\omega_{p4}$) is introduced, with their frequency difference precisely tuned to $\omega_1$-$\omega_2$. This generates an optomechanical parametric coupling between the two phonon modes, which is mapped to an effective beam-splitter Hamiltonian, as shown by the second term in Eq. (1). The complete dynamics of the system, as described in Eq. (1), can be modelled by using the following quantum Langevin equation: $\partial_t \hat{b}_j = -(\gamma_j^{\text{eff}} \hat{b}_j + i\eta \hat{b}_k + \delta_{nj}\Gamma_{\text{par}} \hat{b}_j^\dagger)/2 + \sqrt{\gamma_j^{\text{eff}}}\, \hat{b}_j^{\text{in}}$ $(n = 1 \text{ or } 2)$ and $j,\ k \in \{1,2\},\ j \neq k$ . The quantum noise operator $\hat{b}_j^{\text{in}}$ arises from an equivalent reservoir with an effective temperature $T_j^{\text{eff}}$, which represents a balance between intrinsic thermal bath and the optomechanical cooling mechanism. The statistics are governed by the correlation functions $\left\langle \hat{b}_j^{\text{in}\dagger}(t) \hat{b}_j^{\text{in}}(t') \right\rangle = \overline{n}_j^{\text{eff}} \delta(t - t')$ , with the effective thermal occupancy $\overline{n}_j^{\text{eff}} = (e^{\hbar\omega_j / k_B T_j^{\text{eff}}} - 1)^{-1}$.

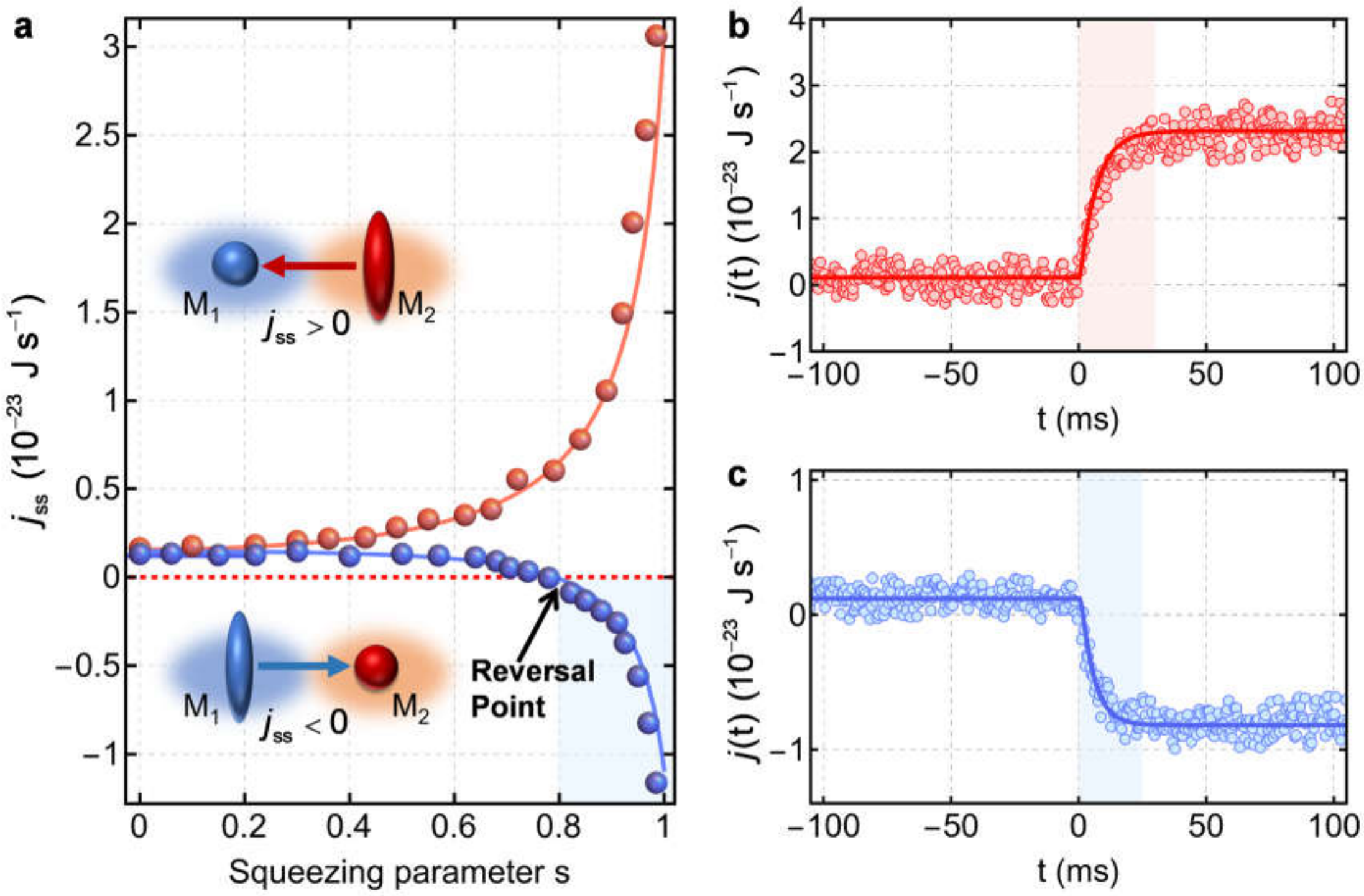

**Fig. 2. Observations of squeezing-engineered phonon heat current. a**, Measured steady-state phonon heat current $j_{ss}$ as a function of the squeezing parameter s, showing heat current amplification (red) and reversal (blue). **b**, **c,** Temporal evolutions of the heat current in the amplification (**b**) and reversal (**c**) regimes, respectively. Each curve represents an experimental measurement obtained by averaging 2500 independent trajectories. The shaded regions highlight the transient processes under the activating phonon squeezing (s=0.95) at t=0 ms. Circles denote experimental data, and solid lines denote simulation results.

To demonstrate on-demand control over phononic thermal transport, the phonon heat current is experimentally measured to characterize the heat exchange between two phonon modes, as illustrated in Fig. 2. The phonon heat currents from the phonon mode $M_2$ to $M_1$ are defined both in the transient state and steady state, as $j(t) = \hbar\omega_2\eta\,\mathrm{Im}\left\langle \hat{b}_1^\dagger(t)\hat{b}_2(t)\right\rangle$ and $j_{ss} = \lim_{t\to\infty} j(t)$, respectively [40].

We first initialize both modes $M_1$ and $M_2$ near their quantum ground states, with effective temperatures $T_1^{eff} = 0.57$ mK and $T_2^{eff} = 2.1$ mK, thereby creating a temperature gradient $\Delta T^{eff} = T_2^{eff} - T_1^{eff}$ across the two nonequilibrium thermal baths. As shown in Fig. 2a, in the absence of squeezing (s=0), a steady-state heat current $j_{ss} = 0.15\times10^{-23}\,\mathrm{J\ s^{-1}}$ emerges due to the optomechanically induced coherent coupling ($\eta/2\pi = 90$Hz).

Applying squeezing to the hotter mode ($M_2$) amplifies the steady-state heat current $j_{ss}$ more than twentyfold as s→1 (red dots in Fig. 2a). This dramatic enhancement arises because antisqueezing intensifies $M_2$'s fluctuations beyond the thermal bath driving. Beyond the critical threshold of s = 1, the system undergoes a dynamical phase transition, marking the onset of the phonon laser regime characterized by coherent self-sustained oscillations [41]. This limitation can be bypassed through feedback control, which facilitates stable dynamics in the s > 1 regime [42]. In contrast, when squeezing is applied to the colder mode $M_1$, $j_{ss}$ decreases toward zero as s increases and eventually reverses direction beyond a reversal point at s ~ 0.8 (blue dots in Fig. 2a). Therefore, this squeezing-engineered phonon heat transport not only amplifies the steady-state heat current, but also effectively overcomes the imposed temperature gradient, enabling anomalous heat flow from the colder to the hotter membrane.

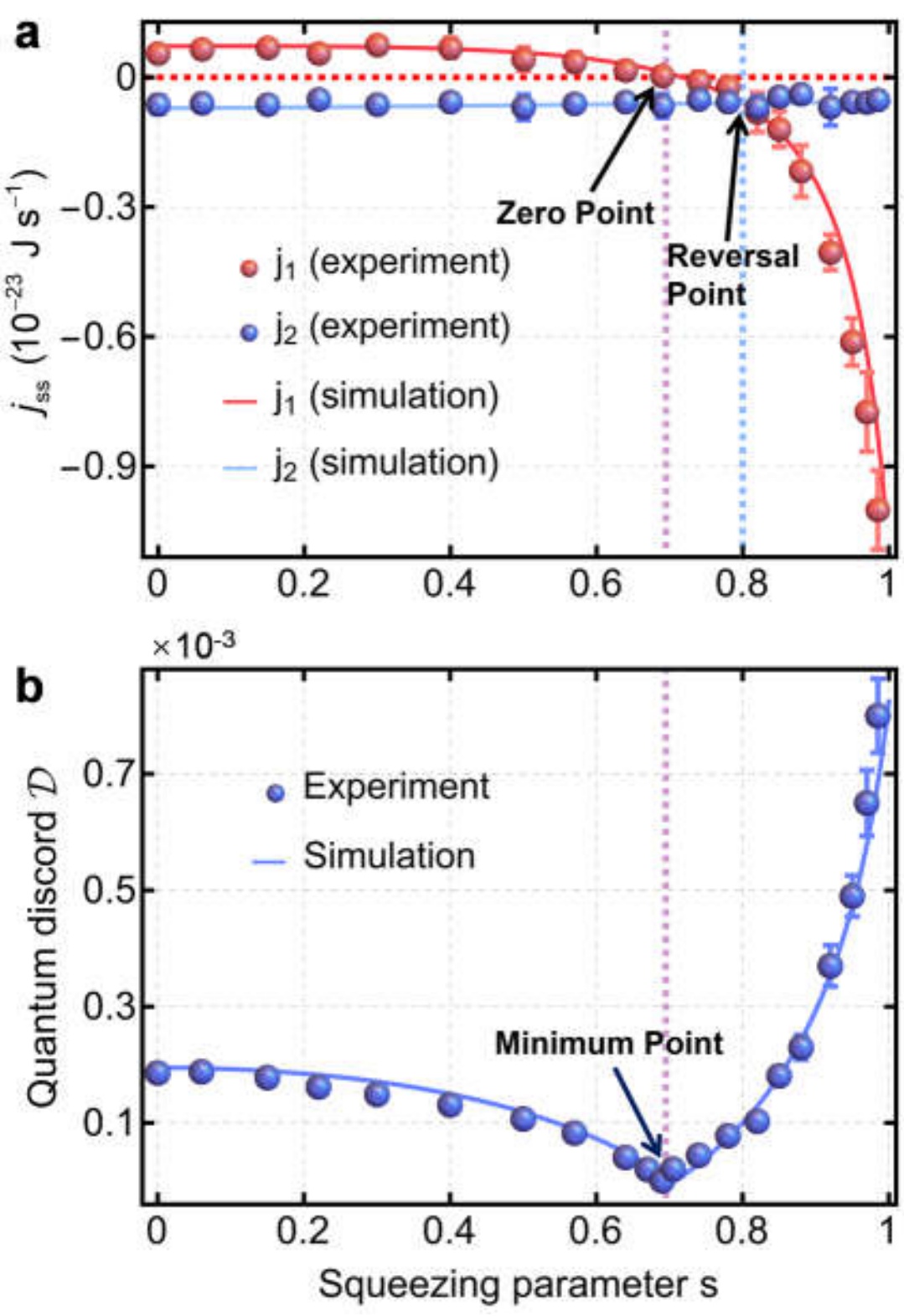

**Fig. 3. Correspondence between heat channels and quantum discord. a**, **b**, Heat channels (**a**) and quantum discord (**b**) of steady-state phonon heat current as a functions of squeezing parameter s. The circles (experiment) and solid lines (simulation). The vertical dashed lines mark the squeezing parameter s where $j_1$ crosses zero (purple), the quantum discord reaches its minimum (purple), and the overall heat current reverses direction (blue).

We further evaluate the dynamic response of thermal transport control. The temporal evolutions of heat current are recorded after instantaneously activating squeezing (s=0.95) at t=0 for both the amplification (Fig. 2b) and reversal (Fig. 2c) regimes. Within the framework of stochastic thermodynamics, the heat current along a single trajectory is inherently subjected to pronounced stochastic fluctuations [43,44]. To capture the underlying transient dynamics, we rapidly switch the squeezing drive between on and off resonance, implementing a periodic quench protocol. This approach allows for the simultaneous real-time acquisition of individual trajectories, enabling the reconstruction of the system's ensemble-averaged behavior from the fluctuating signals [45]. Exponential fitting revels a characteristic response time within 30 milliseconds, exhibiting a switching speed that surpasses recent sub-second heat-flow switching schemes by more than an order of magnitude [14]. This timescale is determined by the effective thermalization rate of the coupled phonon modes and squeezing parameter, which can be further optimized by enhancing the effective mechanical damping rates.

To unveil the microscopic mechanism underlying squeezing-engineered phonon heat transport, we investigate the phonon heat structure, i.e. the heat current components originating from cross-quadrature correlations of mechanical motions, with particular focus on the heat current reversal regime. For a Gaussian system, the steady-state properties of the coupled phononic system are fully encoded in the covariance matrix $\sigma$ with matrix element $\sigma_{ij} = \langle u_i u_j + u_j u_i \rangle / 2$, where $u = (X_1,\ P_1,\ X_2,\ P_2)^{\mathrm{T}}$ represents the vector of dimensionless quadrature operators of the phonon mode $X_{1,2} = (\hat{b}_{1,2} + \hat{b}_{1,2}^{\dagger}) / \sqrt{2}$ and $P_{1,2} = (\hat{b}_{1,2} - \hat{b}_{1,2}^{\dagger}) / i\sqrt{2}$. As illustrated in Fig. 3a, the steady state heat current $j_{ss}$ in the reversal regime (Fig. 2a) arises from the contributions of both heat transport channels from two distinct cross-quadrature correlations, i.e. $\mathrm{j}_1 = -\hbar\omega_2\eta\sigma_{23} / 2$ and $\mathrm{j}_2 = -\hbar\omega_2\eta\sigma_{14} / 2$, such that $j_{ss} = \mathrm{j}_1 - \mathrm{j}_2$, where $\mathrm{j}_1$($-\mathrm{j}_2$) represents the mean power exerted on $M_1$ by $M_2$ via kinetic $P_1P_2$ (potential $X_1X_2$) coupling [46]. In the absence of squeezing (s = 0), both channels exhibit symmetric magnitudes. This degeneracy originates from isotropic phase space fluctuations, a signature of thermodynamic reciprocity protected by the continuous U(1) symmetry. As the squeezing parameter s increases, this reciprocal symmetry is explicitly broken, inducing phase-space anisotropy in $M_1$. While $\mathrm{j}_2$ remains negative and nearly constant (blue dots in Fig. 3a), $\mathrm{j}_1$ decreases from its initial positive value, crosses zero at s ~ 0.7, thereafter reverses sign completely (red dots in Fig. 3a). Heat transport is therefore dominated by the $\mathrm{j}_1$ channel under antisqueezing of the $P_1$ quadrature. The condition $\mathrm{j}_1 = \mathrm{j}_2$ (blue dashed line) yields a reversal point at s ~ 0.8 for the net current $j_{ss}$, quantitatively matching the reversal in Fig. 2a. Consequently, squeezing reconfigures the energy-exchange landscape.

To provide a quantitative understanding of this behavior, we employ the Lyapunov equation for the reversal regime, i.e. with squeezing applied to $M_1$, and derive the following analytical expressions for the two transport channels, $j_1 \propto \left[\left(\bar{n}_2^{\text{eff}}+1/2\right)-\left(\bar{n}_1^{\text{eff}}+1/2\right)/(1-s)\right]$ and $j_2 \propto \left[\left(\bar{n}_1^{\text{eff}}+1/2\right)/(1+s)-\left(\bar{n}_2^{\text{eff}}+1/2\right)\right]$. Here the $j_1$ and $j_2$ are asymmetrically modified by the squeezing-induced factors $(1\text{-s})^{-1}$ and $(1\text{+s})^{-1}$, respectively (see Supplementary Note 6). Interestingly, in the quantum ground-state limit ($\bar{n}_1^{\text{eff}}=0$), the factor $(1\text{-s})^{-1}$ directly amplifies zero-point fluctuations (1/2), indicating that the squeezing drive can harvest vacuum fluctuations to power heat transport. This phenomenon is rooted in the squeezing-induced $Z_2$ anisotropy, characterized by the phase-sensitive redistribution of quadrature fluctuations by factors $(1\pm s)^{-1}$ along the principal axes in squeezed phase space. By breaking the U(1) symmetry, the squeezing enables the system to selectively amplify or suppress specific transport channels depending on the phase alignment. Consequently, the anomalous heat transport arises from a microscopic reconfiguration of cross-correlations via squeezing, rather than a modified effective temperature (see supplementary Note 5 for more discussions).

To further verify the quantum signature of the observed phonon heat transport, we investigate the link between the heat current and quantum discord. Quantum discord ($\mathcal{D}$) quantifies the non-classical correlations of a system by evaluating the discrepancy between the total quantum mutual information and the maximum classical correlation extractable via local measurements [47,48]. As a robust measure of quantumness, quantum discord captures correlations that persist even in the absence of bipartite entanglement. For our two-mode Gaussian system, quantum discord can be efficiently evaluated through the symplectic invariants of the system's covariance matrix (see Supplementary Note 7). As shown in Fig. 3b, the experimentally estimated $\mathcal{D}$ exhibits a minimum at s ~ 0.7, which coincides precisely with the zero crossing of $j_1$ (red dots in Fig. 3a). Both $\mathcal{D}$ and $|j_1|$ reach their minima simultaneously before rising as s increases. Such a synchronous trend is notably absent between $\mathcal{D}$ and $j_{ss}$, indicating that the quantum correlations are more fundamentally coupled to individual transport channels than to the total heat flow. Unlike a previous experiment that demonstrated transient heat reversal by consuming finite, pre-established quantum correlations [49], our results reveal that heat control in both the steady state and transient regime is achieved by actively reconfiguring the phase-space of a nonequilibrium open system via squeezing.

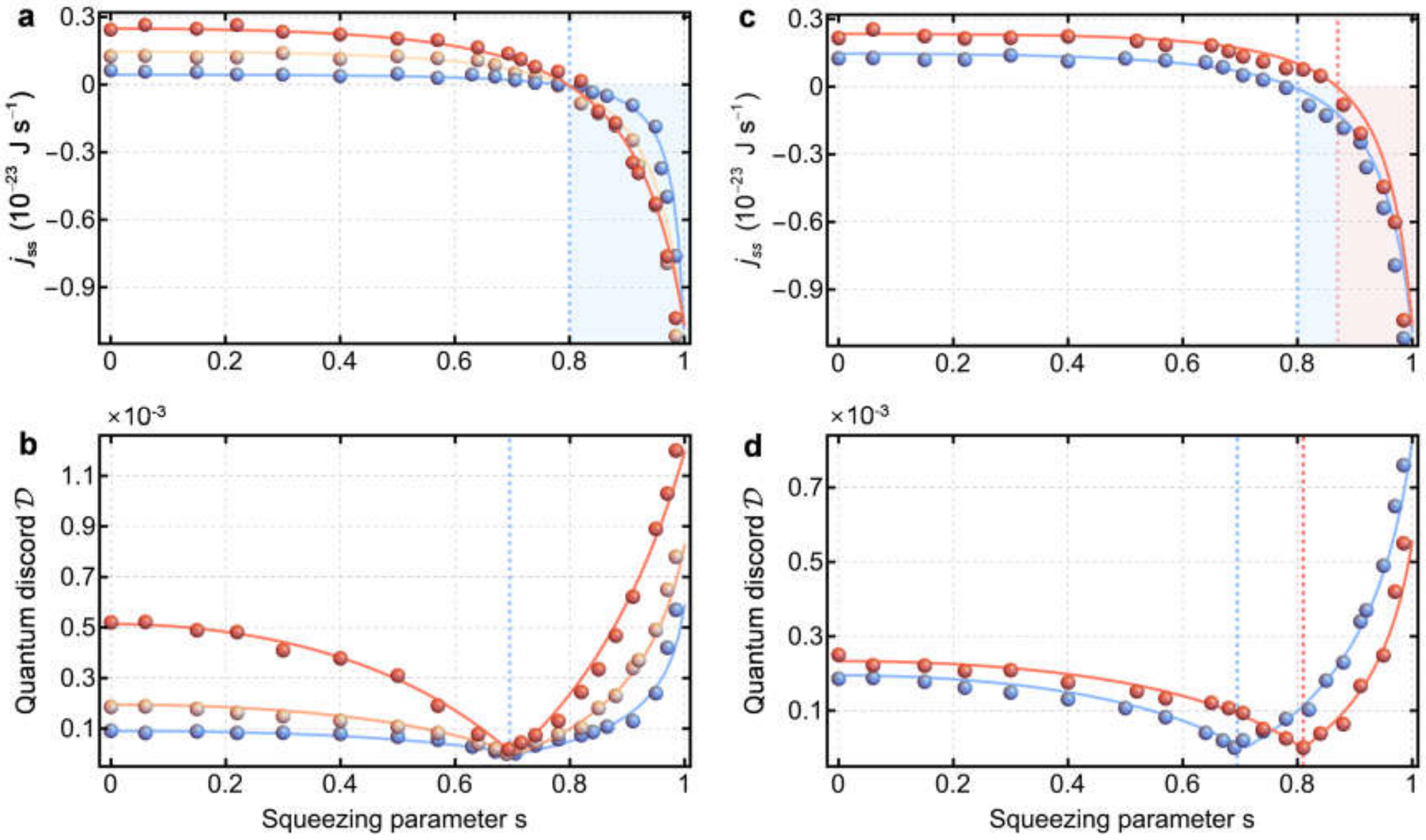

**Fig. 4. Controllability of squeezing-engineered phonon heat transport. a**, **b**, Steady-state heat current $j_{ss}$ (**a**) and the corresponding quantum discord $\mathcal{D}$ (**b**) as functions of squeezing parameter s for coupling strength η/2π = 55 (blue), 90 (yellow), and 145 (red) Hz. **c**, **d**, $j_{ss}$ (**c**) and $\mathcal{D}$ (**d**) as functions of s for $\Delta T^{\rm eff}$ = 1.56 mK (blue) and 2.77 mK (red). The circles (experiment) and solid lines (simulation).

Lastly, we explore the versatility of heat transport control by investigating the dependence of the steady-state heat current $j_{ss}$ and quantum discord $\mathcal{D}$ on the coherent coupling strength η and the thermal gradient $\Delta T^{\rm eff}$. As illustrated in Figs. 4a and 4b, increasing η via the power of the second pair of beating lasers at fixed s enhances both $j_{ss}$ and $\mathcal{D}$ by enabling more efficient heat transfer between the phonon modes. Notably, the reversal point of $j_{ss}$ and the local minimum of $\mathcal{D}$ remain invariant with respect to η. Figures 4c and 4d show that increasing the temperature gradient $\Delta T^{\mathrm{eff}}$ displaces both the reversal point of $j_{\mathrm{ss}}$ and the minimum of $\mathcal{D}$ toward higher values of s, revealing a competition between squeezing and temperature gradient.

In conclusion, we have experimentally demonstrated squeezing-engineered phonon heat transport in a two-membrane cavity optomechanical system. By breaking the inherent phase symmetry of the system via optomechanical squeezing, we show that quantum squeezing can reconfigure heat flux between near-ground-state mechanical resonators, enabling arbitrary real-time control—amplification, reduction, and reversal—of phonon heat transport. Our results establish a fundamental link between steady-state heat current and the emergence of quantum correlations, offering a microscopic picture of anomalous thermal transport, and position quantum squeezing as a versatile, symmetry-breaking platform for on-demand phononic thermal control—a paradigm we anticipate will extend far beyond phonons to enable real-time, programmable manipulation of transport across a wide range of bosonic systems, including photons, magnons, polaritons, and even cold-atom currents [50], thereby opening new frontiers for active quantum thermal management, reconfigurable logic, nonreciprocal energy flow, and ultimately, the exploration of information-energy relations in quantum thermodynamics.

[1] N. Li, J. Ren, L. Wang, G. Zhang, P. Hänggi, and B. Li, Colloquium: Phononics: Manipulating heat flow with electronic analogs and beyond, Rev. Mod. Phys. 84, 1045 (2012).

[2] M. Maldovan, Sound and heat revolutions in phononics, Nature 503, 209 (2013).

[3] G. Wehmeyer, T. Yabuki, C. Monachon, J. Wu, and C. Dames, Thermal diodes, regulators, and switches: Physical mechanisms and potential applications, Appl. Phys. Rev. 4, 041304 (2017).

[4] Y. Li, W. Li, T. Han, X. Zheng, J. Li, B. Li, S. Fan, and C. W. Qiu, Transforming heat transfer with thermal metamaterials and devices, Nat. Rev. Mater. 6, 488 (2021).

[5] Jukka P. Pekola and Bayan Karim, Colloquium: Quantum heat transport in condensed matter systems, Rev. Mod. Phys. 93, 041001 (2021).

[6] M. C. Mackey, Time's Arrow: The Origins of Thermodynamic Behavior (Springer, New York, 1992).

[7] J. M. Ziman, Electrons and Phonons: The Theory of Transport Phenomena in Solids (Oxford Univ. Press, 1960).

[8] B. Li, L. Wang, and G. Casati, Thermal Diode: Rectification of Heat Flux, Phys. Rev. Lett. **93**, 184301 (2004).

[9] C. Chang, D. Okawa, A. Majumdar, and A. Zettl, Solid-state thermal rectifier, Science 314, 1121 (2006).

[10] P. Ben-Abdallah and S.-A. Biehs, Phase-change radiative thermal diode, Appl. Phys. Lett. 103, 191907 (2013).

[11] Y. Li, X. Y. Shen, Z. H. Wu, J. Y. Huang, Y. X. Chen, Y. S. Ni, and J. P. Huang, Temperature-dependent transformation thermotics: from switchable thermal cloaks to macroscopic thermal diodes, Phys. Rev. Lett. 115, 195503 (2015).

[12] Y. Zhang, Q. Lv, H. Wang, S. Zhao, Q. Xiong, R. Lv, and X. Zhang, Simultaneous electrical and thermal rectification in a monolayer lateral heterojunction, Science 378, 169 (2022).

[13] A. Sood, F. Xiong, S. Chen, H. Wang, D. Selli, J. Zhang, C. J. McClellan, J. Sun, D. Donadio, Y. Cui, E. Pop, and K. E. Goodson, An Electrochemical Thermal Transistor, Nat. Commun. 9, 4510 (2018).

[14] J. W. Lim, A. Majumder, R. Mittapally, A.-R. Gutierrez, Y. Luan, E. Meyhofer, and P. Reddy, A nanoscale photonic thermal transistor for sub-second heat flow switching, Nat. Commun. 15, 5584 (2024).

[15] L. Wang and B. Li, Thermal Logic Gates: Computation with Phonons, Phys. Rev. Lett. 99, 177208 (2007).

[16] B. L. Wooten, R. Iguchi, P. Tang, J. S. Kang, K.-i. Uchida, G. E. W. Bauer, and J. P. Heremans, Electric field–dependent phonon spectrum and heat conduction in ferroelectrics, Sci. Adv. 9, eadd7194 (2023).

[17] M. Li, H. Wu, E. M. Avery, Z. Qin, D. P. Goronzy, H. D. Nguyen, T. Liu, P. S. Weiss, and Y. Hu, Electrically gated molecular thermal switch, Science 382, 585

(2023).

[18] C. Liu, Y. Si, H. Zhang, C. Wu, S. Deng, Y. Dong, Y. Li, M. Zhuo, N. Fan, B. Xu, P. Lu, L. Zhang, X. Lin, X. Liu, J. Yang, Z. Luo, S. Das, L. Bellaiche, Y. Chen, and Z. Chen, Low voltage–driven high-performance thermal switching in antiferroelectric PbZrO3 thin films, Science 382, 1265 (2023).

[19] Y. Sheng, H. Zhu, S. Xie, Q. Lv, H. Xie, H. Wang, R. Lv, and H. Bao, Electrically-driven reversible phonon transport manipulation in two-dimensional heterostructures, Nat. Commun. 16, 1970 (2025).

[20] P. Upreti, D. Rashadfar, R. Sahul, D. L. Abernathy, J. P. Heremans, R. P. Hermann, and M. E. Manley, Electric Field Control of Phonon Lifetimes and Thermal Conductivity in Relaxor-Based Ferroelectric, PRX Energy 5, 013001 (2026).

[21] H. Jin, O. D. Restrepo, N. Antolin, S. R. Boona, W. Windl, R. C. Myers, and J. P. Heremans, Phonon-induced diamagnetic force and its effect on the lattice thermal conductivity, Nat. Mater. 14, 601 (2015).

[22] X. Meng, T. Pandey, J. Jeong, S. Fu, J. Yang, K. Chen, A. Singh, F. He, X. Xu, J. Zhou, W.-P. Hsieh, A. K. Singh, J.-F. Lin, and Y. Wang, Thermal conductivity enhancement in under extreme strain, Phys. Rev. Lett. 122, 155901 (2019).

[23] E. Langenberg, D. Saha, M. E. Holtz, J.-J. Wang, D. Bugallo, E. Ferreiro-Vila, H. Paik, I. Hanke, S. Ganschow, D. A. Muller, L.-Q. Chen, G. Catalan, N. Domingo, J. Malen, D. G. Schlom, and F. Rivadulla, Ferroelectric domain walls in are effective regulators of heat flow at room temperature, Nano Lett. 19, 7901 (2019).

[24] J. Shin, J. Sung, M. Kang, X. Xie, B. Lee, K. M. Lee, T. J. White, C. Leal, N. R. Sottos, P. V. Braun, and D. G. Cahill, Light-triggered thermal conductivity switching in azobenzene polymers, Proc. Natl. Acad. Sci. 116, 5973 (2019).

[25] K. Y. Fong, H-K. Li, R. Zhao, S. Yang, Y. Wang, and X. Zhang, Phonon heat transfer across a vacuum through quantum fluctuations, Nature 576, 243 (2019).

[26] C. Yang, X. Wei, J. Sheng, and H. Wu, Phonon heat transport in cavity-mediated optomechanical nanoresonators, Nat. Commun. 11, 4656 (2020).

[27] J. del Pino, J. J. Slim, and E. Verhagen, Non-Hermitian chiral phononics through optomechanically induced squeezing, Nature 606, 82 (2022).

[28] L. Tang, J. Tang, M. Chen, F. Nori, M. Xiao, and K. Xia, Quantum squeezing induced optical nonreciprocity, Phys. Rev. Lett. 128, 083604 (2022).

[29] T.-X. Lu, Y. Wang, K. Xia, X. Xiao, L.-M. Kuang, and H. Jing, Quantum squeezing induced nonreciprocal phonon laser, Sci. China-Phys. Mech. Astron. 67, 260312 (2024).

[30] S. Sonar, M. Hajdusek, M. Mukherjee, R. Fazio, V. Vedral, S. Vinjanampathy, and L.-C. Kwek, Squeezing Enhances Quantum Synchronization, Phys. Rev. Lett. 120, 163601 (2018).

[31] C. J. Zhu, L. L. Ping, Y. P. Yang, and G. S. Agarwal, Squeezed Light Induced Symmetry Breaking Superradiant Phase Transition, Phys. Rev. Lett. 124, 073602

(2020).
[32] S. C. Burd, R. Srinivas, H. M. Knaack, W. Ge, A. C. Wilson, D. J. Wineland, D. Leibfried, J. J. Bollinger, D. T. C. Allcock, and D. H. Slichter, Quantum amplification of boson-mediated interactions, Nat. Phys. 17, 898–902 (2021).
[33] G.-L. Zhu, C.-S. Hu, H. Wang, W. Qin, X.-Y. Lv, and F. Nori, Nonreciprocal superradiant phase transitions and multicriticality in a cavity QED system, Phys. Rev. Lett. 132, 193602 (2024).
[34] X. Wei, J. Sheng, C. Yang, Y. Wu, and H. Wu, Controllable two-membrane-in-the-middle cavity optomechanical system, Phys. Rev. A 99, 023851 (2019).
[35] P. Piergentili, L. Catalini, M. Bawaj, S. Zippilli, N. Malossi, R. Natali, D. Vitali, and G. Di Giuseppe, Two-membrane cavity optomechanics, New J. Phys. 20, 083024 (2018).
[36] X. Yao, M. H. J. De Jong, J. Li, and S. Gröblacher, Long-range optomechanical interactions in SiN membrane arrays, Phys. Rev. X 15, 011014 (2025).
[37] R. W. Peterson, T. P. Purdy, N. S. Kampel, R. W. Andrews, P.-L. Yu, K. W. Lehnert, and C. A. Regal, Laser cooling of a micromechanical membrane to the quantum backaction limit, Phys. Rev. Lett. 116, 063601 (2016).
[38] A. Chowdhury, P. Vezio, M. Bonaldi, A. Borrielli, F. Marino, B. Morana, G. A. Prodi, P. M. Sarro, E. Serra, and F. Marin, Quantum Signature of a Squeezed Mechanical Oscillator, Phys. Rev. Lett. 124, 023601 (2020).
[39] S. Wu, J. Sheng, X. Zhang, Y. Wu, and H. Wu, Parametric excitation of a SiN membrane via piezoelectricity, AIP Adv. 8, 015209 (2018).
[40] S.-A. Biehs and G. S. Agarwal, Dynamical quantum theory of heat transfer between plasmonic nanosystems, J. Opt. Soc. Am. B 30, 700 (2013).
[41] D. Bothner, S. Yanai, A. Iniguez-Rabago, M. Yuan, Y. M. Blanter, and G. A. Steele, Cavity electromechanics with parametric mechanical driving, Nat. Commun. 11, 1589 (2020).
[42] M. Poot, K. Y. Fong, and H. X. Tang, Deep feedback-stabilized parametric squeezing in an opto-electromechanical system, New J. Phys. 17, 043056 (2015).
[43] Udo Seifert, Stochastic thermodynamics, fluctuation theorems and molecular machines, Rep. Prog. Phys. 75, 126001 (2012).
[44] S. Ciliberto, Experiments in Stochastic Thermodynamics: Short History and Perspectives, Phys. Rev. X 7, 021051 (2017).
[45] C. Yang, N. Deng, J. Sheng, and H. Wu, Entropy measurement of nonequilibrium phonon heat transport, Commun. Phys. 8, 289 (2025).
[46] Gabriel Barton, Classical van der Waals heat flow between oscillators and between half-spaces, J. Phys.: Condens. Matter 27, 214005 (2015).
[47] G. Adesso and A. Datta, Quantum versus classical correlations in Gaussian states, Phys. Rev. Lett. 105, 030501 (2010).

[48] P. Giorda and M. G. A. Paris, Gaussian quantum discord, Phys. Rev. Lett. 105, 020503 (2010).
[49] K. Micadei, J. P. S. Peterson, A. M. Souza, R. S. Sarthour, I. S. Oliveira, G. T. Landi, T. B. Batalhão, R. M. Serra, and E. Lutz, Reversing the direction of heat flow using quantum correlations, Nat. Commun. 10, 2456 (2019).
[50] P. Fabritius, J. Mohan, M. Talebi, S. Wili, W. Zwerger, M.-Z. Huang, and T. Esslinger, Irreversible entropy transport enhanced by fermionic superfluidity, Nat. Phys. 20, 1091 (2024).